\documentstyle [12pt,axodraw,epsf] {article}

\catcode`@=12
\newcommand{\beq}{\begin{equation}}
\newcommand{\eeq}{\end{equation}}
\newcommand{\nn}{\nonumber}
\newcommand{\bea}{\begin{eqnarray}}
\newcommand{\eea}{\end{eqnarray}}

\newcommand{\rfn}[1]{(\ref{#1})}
\newcommand{\Eq}[1]{Eq.~(\ref{#1})}

\newcommand{\ea}{{\it et al.}}

\newcommand{\plet}[1]{{ Phys. Lett. }{\bf #1}}
\newcommand{\pr}[1]{{ Phys. Rev. }{\bf #1}}
\newcommand{\prlet}[1]{{ Phys. Rev. Lett. }{\bf #1}}

\newcommand{\epj}[1]{{ Eur. Phys. J. }{\bf #1}}

\def\lsim{\mathrel{\vcenter{\hbox{$<$}\nointerlineskip\hbox{$\sim$}}}}

\begin{document}
\thispagestyle{empty}
\begin{flushright} CPT-2000/P.4083\\UCRHEP-T293\\ November 2000\
\end{flushright}
\vspace{0.5in}
\begin{center}
{\Large	\bf Allowable Low-Energy E$_6$ Subgroups from Leptogenesis\\}
\vspace{1.3in}
{\bf Thomas Hambye$^{1,2}$, Ernest Ma$^3$, Martti Raidal$^{3,4}$, and Utpal 
Sarkar$^5$\\}
\vspace{0.2in}
{\sl $^1$ Centre de Physique Theorique, CNRS Luminy, Marseille 13288, France\\}
{\sl $^2$ INFN - Laboratori Nazionali di Frascati, 00044 Frascati, Italy\\}
{\sl $^3$ Physics Department, University of California, Riverside, 
California 92521, USA\\}
{\sl $^4$ National Institute of Chemical and Biological Physics, 10143 
Tallinn, Estonia\\}
{\sl $^5$ Physical Research Laboratory, Ahmedabad 380 009, India\\}
\vspace{1.3in}
\end{center}
\begin{abstract}\
There are only two viable low-energy $E_6$ subgroups: 
$SU(3)_C \times SU(2)_L \times U(1)_Y \times U(1)_N$ or $SU(3)_C \times 
SU(2)_L \times SU(2)'_R \times U(1)_{Y_L + Y'_R}$, which would not erase any 
preexisting lepton asymmetry of the Universe that may have been created by 
the decay of heavy singlet (right-handed) neutrinos or any other 
mechanism.  They are also the two most favored 
$E_6$ subgroups from a recent analysis of present neutral-current data. 
We study details of the leptogenesis, as well as some salient 
experimental signatures of the two models.
\end{abstract}

\newpage
\baselineskip 24pt

In the energy range of 100 GeV to 1 TeV, physics beyond the standard model 
(SM) may appear in two ways.  One is the possible addition of supersymmetry;
the other is the possible extension of the  $SU(3)_C \times SU(2)_L 
\times U(1)_Y$ gauge group to a larger symmetry group $G$.  
Both of these options are realized in the  $E_6$ superstring models 
which predict the existence of new particles,
such as an extra gauge boson $Z',$ at ${\cal O}(1)$ TeV \cite{hewett}.

As required by the solar and atmospheric neutrino data \cite{neutrino},
any extension of the SM should 
include a mechanism for generating small nonzero neutrino masses.
It should also be consistent with the present 
observed baryon asymmetry of the Universe.  If it contains $B-L$ violating 
interactions at energy scales in the range $10^2-10^{12}$ GeV, 
these together with the $B+L$ violating electroweak sphalerons \cite{kurush} 
would erase \cite{erase} whatever lepton or baryon asymmetry that may have 
been created at an earlier epoch of the Universe \cite{ft}.

In this Letter we show that if $G$ is a 
subgroup of $E_6$, and if $G$ survives down to ${\cal O}(1)$ TeV as is
expected in these theories,
then the  constraint of successful leptogenesis \cite{fuya,buchmuller} 
from the decay of heavy singlet (right-handed) neutrinos $N$ results 
{\it uniquely} in only two possible candidates.  
One is $G_1=SU(3)_C \times SU(2)_L \times U(1)_Y \times U(1)_N$ \cite{ma1}, 
and the other is 
$G_2=SU(3)_C \times SU(2)_L \times SU(2)'_R \times U(1)_{Y_L+Y'_R}$ 
\cite{ma2}, where $SU(2)'_R$ is not the conventional $SU(2)_R$.
Only these groups allow $N$ to have zero quantum numbers with respect to 
all of their transformations. 
Any other subgroup of $E_6$  would result in lepton-number violating 
interactions at ${\cal O}(1)$ TeV as it is broken down to the SM. 
Remarkably, $G_{1,2}$ happen to be also the two most favored $E_6$ 
subgroups from a recent analysis \cite{erla} of present neutral-current data.  
This is a possible hint that one of these two models may in fact be correct. 

Whereas there is only one version \cite{ma2} of the model based on $G_2$, 
we find 2 (and only 2) phenomenologically viable versions of $G_1$, 
and work out the details of the leptogenesis in all 3 cases. 
In addition to specific $Z'$ properties at colliders, 
we also predict the discovery of $W_R^\pm$ in the $G_2$ model. 
Among other distinctive experimental signatures are the s-channel 
diquark resonances at hadron colliders, which can be tested up to the 
{\it multi} TeV scale at the LHC \cite{res}.

The fundamental \underline {27} representation of $E_6$ may be classified 
according to its maximal subgroup $SU(3)_C \times SU(3)_L \times SU(3)_R.$ 
In the notation where all fermions are considered left-handed, one has
the particle assignment
\begin{equation}
(u,d,h) \sim (3,3,1), ~~~ (h^c, d^c, u^c) \sim (3^*,1,3^*),
\end{equation}
whereas $\nu_e,e,e^c$ together with the new superfields 
$N^c,\nu_E,E,N^c_E,E^c,S$ are 
contained in $(1,3^*,3).$ 
Under the decomposition 
$SU(3)_L\to SU(2)_L \times U(1)_{Y_L},$ 
$SU(3)_R\to SU(2)_R \times U(1)_{Y_R},$
they may be represented pictorically as
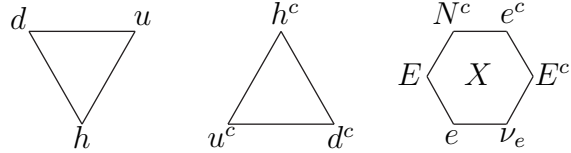
\begin{center}
\begin{picture}(200,50)(0,5)
\Line(5,45)(45,45)
\Line(5,45)(25,10)
\Line(45,45)(25,10)
\Text(1,50)[]{$d$}
\Text(48,50)[]{$u$}
\Text(25,5)[]{$h$}
\Line(80,10)(120,10)
\Line(80,10)(100,45)
\Line(120,10)(100,45)
\Text(123,5)[]{$d^c$}
\Text(78,5)[]{$u^c$}
\Text(102,52)[]{$h^c$}
\Line(165,45)(185,45)
\Line(165,10)(185,10)
\Line(165,45)(155,28)
\Line(185,10)(195,28)
\Line(185,45)(195,28)
\Line(165,10)(155,28)
\Text(165,5)[]{$e$}
\Text(188,5)[]{$\nu_e$}
\Text(165,52)[]{$N^c$}
\Text(188,52)[]{$e^c$}
\Text(149,29)[]{$E$}
\Text(203,29)[]{$E^c$}
\Text(175,29)[]{$X$}
\end{picture}
\end{center}
where 
the horizontal axis measures $T_{3L} + T_{3R},$ the vertical 
axis $Y_L + Y_R$, and $X=\nu_E,S,N_E^c.$
In this particle assignment, the assumption is 
that the $SU(2)_R$ subgroup contains the quark doublet 
$(d^c,u^c)$ as in the usual left-right model.  However, as was first 
pointed out in Ref.~\cite{ma2}, a different decomposition of $SU(3)_R$ 
may be chosen, i.e. $SU(2)'_R$, where $(h^c,u^c)$ is the doublet.  A third 
way is to choose the direction of symmetry breaking so that $(h^c,d^c)$ is 
a doublet \cite{loro}.  These 3 choices are merely the familiar old $T,V,U$ 
isospins of $SU(3)$.  With the interchange  $d^c\leftrightarrow h^c$ 
in going from $SU(2)_R$ to $SU(2)'_R$, one must also interchange 
$(\nu_e,e)\leftrightarrow (\nu_E,E)$, and $N^c\leftrightarrow S$. 
The new pictorial representation is 
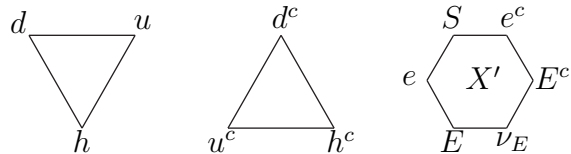
\begin{center}
\begin{picture}(200,50)(0,5)
\Line(5,45)(45,45)
\Line(5,45)(25,10)
\Line(45,45)(25,10)
\Text(1,50)[]{$d$}
\Text(48,50)[]{$u$}
\Text(25,5)[]{$h$}
\Line(80,10)(120,10)
\Line(80,10)(100,45)
\Line(120,10)(100,45)
\Text(123,5)[]{$h^c$}
\Text(78,5)[]{$u^c$}
\Text(102,52)[]{$d^c$}
\Line(165,45)(185,45)
\Line(165,10)(185,10)
\Line(165,45)(155,28)
\Line(185,10)(195,28)
\Line(185,45)(195,28)
\Line(165,10)(155,28)
\Text(165,5)[]{$E$}
\Text(188,5)[]{$\nu_E$}
\Text(165,52)[]{$S$}
\Text(188,52)[]{$e^c$}
\Text(149,29)[]{$e$}
\Text(203,29)[]{$E^c$}
\Text(177,29)[]{$X'$}
\end{picture}
\end{center}
where $X'=\nu_e,N^c,N_E^c.$  
The electric charge is given by
\begin{equation}
Q = T_{3L} + Y, ~~~ Y = Y_L + T_{3R} + Y_R.
\end{equation}
If $SU(2)_R \times U(1)_{Y_R}$ is replaced by $SU(2)'_R \times U(1)_{Y'_R}$, 
then
\begin{equation}
T'_{3R} = {1 \over 2} T_{3R} + {3 \over 2} Y_R, ~~~ Y'_R = {1 \over 2} T_{3R} 
- {1 \over 2} Y_R.
\label{t3r}
\end{equation}
Hence $T'_{3R} + Y'_R = T_{3R} + Y_R$ so 
that $Y$ remains the same as it must.  As far as the SM is concerned, 
the two extensions are equally
viable and no interaction involving only the SM particles, i.e. 
$u,d,u^c,d^c,\nu_e,e,e^c$ and the corresponding gauge bosons, can tell them 
apart. 

Another way to extend the  SM is to attach an extra $U(1).$  In this 
case, $E_6$ offers the choice of a linear combination of two distinct $U(1)$ 
subgroups \cite{badewh}, i.e. $E_6 \to SO(10) \times U(1)_\psi$ and 
$SO(10) \to SU(5) \times U(1)_\chi,$ \nolinebreak  with
\begin{equation}
Q_\psi = \sqrt {3 \over 2} (Y_L - Y_R), ~~~ Q_\chi = \sqrt {1 \over 10} 
(5 T_{3R} - 3 Y).
\end{equation}
Let $Q_\alpha \equiv Q_\psi \cos \alpha + Q_\chi \sin \alpha$, then all 
possible $U(1)$ extensions of the SM under $E_6$ may be studied 
\cite{chhaum} as a function of $\alpha$.  

Let us now discuss the role of $B-L$ in $E_6$ models.
It is well-known that $Y_L + Y_R = (B-L)/2$ as far as the SM 
particles are concerned \cite{pasaso}.  
For the new fermions belonging to the $E_6$ 
fundamental representation, this may be extended as a definition because 
their Yukawa interactions with the SM particles must be 
invariant under $G$. With this assignment, all Yukawa and gauge interactions
{\it conserve} $B-L$. 
Among the five neutral fermions in $E_6$,
only two ($\nu_e$ and $N^c$) carry nonzero $B-L$ quantum
numbers ($-1$ and 1).  Hence 
the {\it only} useful source of $B-L$ violation in any $E_6$ 
model is the large Majorana mass  of $N^c$, which is of course also 
the reason why neutrino masses are small (from the seesaw mechanism) to 
begin with.

In a successful scenario of leptogenesis \cite{fuya}, the decay of 
the {\it physical} heavy Majorana neutrino $N$ (i.e. $N^c$ plus its conjugate) 
must satisfy the out-of-equilibrium condition
\begin{equation}
\Gamma_{N} < H(T=m_{N})=\sqrt{\frac{4 \pi^3 g_{\ast}}{45}} \frac{T^2}{M_P},
\label{noneq}
\end{equation}
where $\Gamma_{N}$ is its decay width, $H(T)$ the Hubble 
expansion rate and $g_\ast$ the effective number of massless degrees of 
freedom at 
the temperature $T$. This requires $m_{N}$ to be many 
orders of magnitude greater than 1 TeV, 
so $N^c$ cannot transform under the low-energy gauge group $G.$
Since $N^c \sim (1;0;-1/2;1/2)$ under $SU(3)_C \times 
T_{3L} \times T_{3R} \times (Y_L+Y_R)$, this group (i.e. the conventional 
left-right model) is forbidden by 
leptogenesis.  On the other hand, $N^c \sim (1;0;0;0)$ under 
$SU(3)_C \times T_{3L} \times T'_{3R} \times (Y_L+Y'_R)$, hence the skew 
left-right model \cite{ma2} is allowed. In the $U(1)_\alpha$ models, $N^c$ 
transforms trivially only if $\tan \alpha = \sqrt{1/15}$.  This is called 
$U(1)_N$ 
\cite{ma1} with
\begin{equation}
Q_N = \sqrt {1 \over 40} (6 Y_L + T_{3R} - 9 Y_R),
\label{qn}
\end{equation}
and is indeed zero for $N^c$, i.e. $Y_L = 1/3$, $T_{3R} = -1/2$, and 
$Y_R = 1/6$.

Thus the only possible $E_6$ subgroups allowed by leptogenesis
are those given by the skew  $SU(2)'_R$ and $U(1)_N$ models.
While details of the leptogenesis and the low-energy phenomenology 
are different in these two models, their choice follows
from a {\it single} and {\it unique} group-theoretical argument 
which has nothing to do with model building. 
Indeed, if not for the fact that $\sin^2 \theta_W \neq 3/8$ at low energies, 
the breaking of $SU(2)'_R \times U(1)_{Y_L+Y'_R}$ would result in 
$U(1)_N \times U(1)_Y$.

There are many virtues \cite{ma1,ma3} associated with these two models. 
They are also the most favored \cite{erla} of 
all known gauge extensions of the SM, based on present 
neutral-current data from atomic parity violation \cite{bewi} and precision 
measurements of the $Z$ width.  The $U(1)_N$ model was not considered in 
Ref.\cite{erla}, but it can easily be included in their Fig.~1 by noting 
that it has $\alpha = 0$ and $\tan \beta = \sqrt {15}$ in their notation, 
thus placing it within the 1$\sigma$ contour together with the $SU(2)'_R$
model.

We shall now work out details of the leptogenesis in these models.
The most general superpotential for the $U(1)_N$ model
coming from the
\underline {27} $\times$ \underline {27} $\times$ \underline {27}
decomposition of the $E_6$ fundamental representation is
\bea
W&=& \lambda_1^{ijk}u^c_i Q_j H^c_k + \lambda_2^{ijk}d^c_i Q_j H_k +
\lambda_3^{ijk}e^c_i L_j H_k  + \label{w} \\
&& \lambda_4^{ijk} S_i  H_j H^c_k + \lambda_5^{ijk} S_i  h_j h^c_k + 
\lambda_6^{ijk}e^c_i h_j u^c_k + \lambda_7^{ijk}h^c_i Q_j L_k + \nn\\ 
&& \lambda_8^{ijk} d^c_i h_j N^c_k + \lambda_9^{ijk}h_i Q_j Q_k + 
\lambda_{10}^{ijk}u^c_i d^c_j h^c_k +\lambda_{11}^{ijk}L_i H^c_j N^c_k, \nn
\eea
where we denote $(\nu_E,E)$ as $H$ and  $(E^c,N^c_E)$ as $H^c$.
The terms $\lambda_{1-5}$ give masses to all fermions and must be present
in any model. $SU(2)_L \times U(1)_Y$ is broken 
by $\langle \tilde{N}_E^c \rangle$ and $\langle \tilde{\nu}_E \rangle,$ 
while $\langle \tilde S \rangle$ breaks $U(1)_N$ [as well as $SU(2)'_R 
\times U(1)_{Y_L + Y'_R}$] and gives masses of order $M_{Z'}$ to
$E$, $h$, $\nu_E$, and $N_E^c$.
Whereas $W$ conserves $B-L$ automatically, there are some terms which 
violate $B+L.$  To prevent 
rapid proton decay, an appropriate $Z_2$ symmetry (extension of R-parity) 
must be imposed. There are 8 ways to do that, resulting in 
8 different models \cite{ma4}. However, the requirements of leptogenesis and
nonzero neutrino masses single out only 2 allowed possibilities.
If $(L,e^c),$ $N^c$ and $(h,h^c)$ are all odd under $Z_2$, then 
$\lambda_{9,10}=0$ in \Eq{w} which is called Model 1. 
Here $N^c$ is a lepton $(L=-1)$ and $h$ is a leptoquark $(B=1/3,$
$L=1).$  If $(h,h^c)$ is even and the others remain odd, then we get Model 2
with $\lambda_{6,7,8}=0$ and $h$ is now a diquark $(B=-2/3).$ 
Note that leptogenesis is also possible in Model 7 of Ref.\cite{ma4} 
with $\lambda_{6-10}=0,$ but as $h$ is stable in this case, it
is ruled out by cosmological considerations. Baryogenesis is also allowed in 
Model 5 of Ref.\cite{ma4} with $\lambda_{6,7,11}=0,$ but since $N^c$ 
is now a baryon with $B=1$ and $L=0$, neutrinos are exactly massless in 
that model.

%
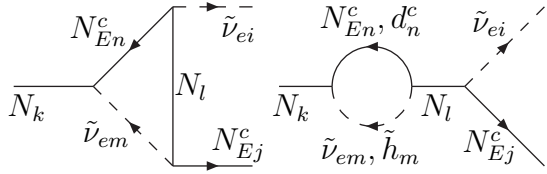
\begin{figure}[t]
\begin{center}
\begin{picture}(200,60)(0,0)
\Line(0,30)(30,30)
\DashArrowLine(60,60)(90,60){5}
\ArrowLine(60,0)(90,0)
\Line(60,60)(60,0)
\ArrowLine(60,60)(30,30)
\DashArrowLine(60,0)(30,30){5}
\Text(5,22)[]{$N_k$}
\Text(35,12)[]{$\tilde\nu_{em}$}
\Text(32,51)[]{$ N^c_{En}$}
\Text(67,30)[]{$N_l$}
\Text(85,52)[]{$\tilde\nu_{ei}$}
\Text(85,8)[]{$ N^c_{Ej}$}
\Line(100,30)(120,30)
\DashArrowArcn(135,30)(15,0,180){5}
\ArrowArc(135,30)(15,0,180)
\Line(150,30)(170,30)
\DashArrowLine(170,30)(200,60){5}
\ArrowLine(170,30)(200,0)
\Text(105,22)[]{$N_k$}
\Text(135,7)[]{$\tilde\nu_{em},\tilde h_m$}
\Text(135,55)[]{$ N^c_{En},d^c_n$}
\Text(160,22)[]{$N_l$}
\Text(180,54)[]{$\tilde\nu_{ei}$}
\Text(180,7)[]{$ N^c_{Ej}$}
\end{picture}
\end{center}
\caption{Loop diagrams interfering with the  $N_k$ tree decay.}
\label{fig}
\end{figure}
The superpotential of the skew $SU(2)'_R$ model is completely fixed 
and can be obtained from \Eq{w}
by setting $\lambda_4=\lambda_3,$ $\lambda_6=-\lambda_5,$ 
$\lambda_7=-\lambda_1$ and $\lambda_{9,10}=0.$ 
Here $h$ is a leptoquark as in the $U(1)_N$
Model 1. However, the $SU(2)'_R$ decomposition also implies that $W_R^-$ 
has $L = 1$ and is {\it odd} under R-parity rather than even.  
Indeed, $W_R^-$ has $T'_{3R} = -1$ and $Y'_R = 0$, but because of 
\Eq{t3r}, it has $Y_R = -1/2.$

In general, the  heavy Majorana neutrino $N_k$ 
decays to the $B-L=-1$ final states $\nu_{e_i} \tilde{N}_{Ej}^c$,
$\tilde{\nu}_{e_i} N_{Ej}^c$, $e_i \tilde{E}^c_j$, $\tilde{e}_i E^c_j$
and $d^c_i \tilde{h}_j$, $\tilde{d}^c_i h_j$ via the interaction terms 
$\lambda_{11}$ and $\lambda_{8}$ in \Eq{w}, respectively, and to their
conjugate states with $B-L=1.$
To establish a $B-L$ asymmetry, one needs:
$(i)$ $B-L$ violation, from the $N$ Majorana mass; 
$(ii)$ CP violation, from the complex couplings $\lambda_{8,11};$ and 
$(iii)$ the out-of-equilibrium condition of \Eq{noneq}.
An equal asymmetry is also generated from the corresponding decays
of the scalar partners $\tilde{N}_k$ \cite{buchmuller}. 
The subsequent decays of  $N^c_{Ej}$, $E^c_j$ and $h_j$ 
or their superpartners to SM particles do not affect the asymmetry 
because they conserve $B-L$.

Technically, the $B-L$ asymmetry $\varepsilon_k$ is generated 
from the interference between tree-level $N_k$ decays and 
one-loop diagrams, some of which are depicted in Fig.\ref{fig} for one 
particular final state. Thus $\varepsilon_k=\varepsilon_k^V+\varepsilon_k^S,$
where $\varepsilon_k^V$ and $\varepsilon_k^S$ are vertex
and self-energy contributions respectively. They are given by
\begin{eqnarray}
\varepsilon_V^k &=& -\frac{1}{8 \pi} \sum_{l,m,n} 
\frac{\sum_{a;i,j}  C_a \mbox{Im}[
\lambda_{a}^{ijk\ast} \lambda_{a}^{mnk\ast} 
\lambda_{a}^{mjl} \lambda_{a}^{inl}
]}
{ \sum_{a;ij} C_a |\lambda_{a}^{ijk}|^2  } 
\nn \\
& &
\sqrt{x_l} \Big[  (1+x_l) \mbox{Log} (1+1/x_l) -1 \Big] , 
\label{epsV} \\
\varepsilon_S^k &=& -\frac{1}{4 \pi} \sum_{l,m,n}  
\frac{\sum_{a,b;i,j} C_{a,b} \mbox{Im}[
\lambda_{a}^{ijk\ast}\lambda_{b}^{mnk\ast} 
\lambda_{b}^{mnl} \lambda_{a}^{ijl}
 ]}
{ \sum_{a;ij} C_a |\lambda_{a}^{ijk}|^2  } 
\nn \\
& & 
\sqrt{x_l} (x_l^2 -1)^{-1} ,
\label{epsS}
\end{eqnarray}
where $x_l=(m_{N_l} / m_{N_k})^2,$ 
indices $a,b=8,11$ denote the interactions of \Eq{w}, and the constants
$C_8=1,$ $C_{11}=2,$ $C_{8,8}=1/2,$ $C_{11,11}=2,$
$C_{8,11}=C_{11,8}=1$ come from the number of diagrams in each case.

Notice two differences from the standard Fukugita-Yanagida 
mechanism \cite{fuya}: 
$(i)$ The structures of the flavor indices in $\varepsilon_k^V$ and 
$\varepsilon_k^S$ are not the same unless there is only one generation of 
scalars.
$(ii)$ There are more self-energy diagrams because the particles
in the loop need not be related to those in the final state. 
This is reflected in \Eq{epsS} by terms which mix the $\lambda_{8}$ and
$\lambda_{11}$ couplings.
Together, $(i)$ and $(ii)$ imply that in contrast to 
the models of Refs. \cite{fuya,buchmuller,fps},
the vertex and self-energy contributions to $\varepsilon_k$ are
{\it not} related to each other, allowing one or the other to be dominant 
independently of the values of the $N_k$ masses. 
This is true even in the $U(1)_N$ Model 2 in which $\lambda_{8}=0.$
Also, in the $U(1)_N$ Model 1 and the $SU(2)'_R$ model, the ordinary 
neutrino masses (induced by $\lambda_{11}LH^cN^c$) need not be related to 
the lepton asymmetry.

The total decay width of $N_k$ is given by 
\begin{equation}
\Gamma_{N_k}= \frac{1}{4 \pi} \sum_{i,j} \left(
| \lambda_8^{ijk} |^2 + 2|\lambda_{11}^{ijk} |^2  \right) m_{N_k} .
\end{equation}
Taking $g_\ast \sim 10^2,$ the out-of-equilibrium condition 
\rfn{noneq} implies
$\sum_{i,j}( |\lambda_8^{ijk} |^2 + 2|\lambda_{11}^{ijk} |^2 ) \lsim  
2 \times 10^{-17}$ GeV$^{-1}$ $m_{N_k}$. 
For $m_{N_k} \sim 10^{15}$ GeV, this gives for example 
$\lambda_8^{ijk}, \lambda_{11}^{ijk} \lsim 10^{-1}$.
As long as  \Eq{noneq} is satisfied, there are no damping effects 
due to the inverse decay or scattering processes which may affect the 
$B-L$ asymmetry.
The baryon-to-entropy ratio generated by the decays 
of $N_k$ and $\tilde{N}_k$ is then
$n_B/s \sim 2 \varepsilon_k n_\gamma/(2 s)=
(\varepsilon_k/g_\ast)(45/\pi^4)$ where $n_\gamma$ is the photon number 
density per comoving volume. In order to satisfy the observed 
value $n_B/s \sim 10^{-10}$, we need $\lambda_{8,11}^{ijk}$ typically of order
$\sim 10^{-3}$ assuming a maximal CP-violating phase. The out-of-equilibrium 
condition can therefore be satisfied easily and the  asymmetry is 
produced with the right order of magnitude.

Above the electroweak phase transition, rapid $B+L$ violating 
sphaleron processes convert the created $B-L$ asymmetry to the 
observed asymmetry of quarks and leptons.
Since the new particle masses are ${\cal O}(1)$ TeV, 
they do not take part in the sphaleron-induced processes.
(Although the anomaly is independent of the masses of the new particles, 
their participation in the sphaleron processes is forbidden by
the phase space available at the time of the electroweak phase transition).
Thus $B$ and $L$ violations in the sphaleron environment 
remain approximately the same as in the SM \cite{harvey}. This completes the 
successful baryogenesis in our models.


There are some unique experimental signatures of the $SU(3)_C \times SU(2)_L 
\times U(1)_Y \times U(1)_N$ and $SU(3)_C \times SU(2)_L \times SU(2)'_R 
\times U(1)_{Y_L + Y'_R}$ models. First, the $Z'$ couplings
are given by $Q_N$ in the former, and by \cite{bahema}
\bea
{-1 \over \sqrt 
{1-2s_w^2}} \left[ s_w^2 Y_L + \left( {3s_w^2-1 \over 2} \right) 
T_{3R} - \left( {3-5s_w^2 \over 2} \right) Y_R \right], \nn
\eea
in the latter. Here $s_w^2 \equiv \sin^2 \theta_W$, assuming  $g_L = g_R$.
For $s_w^2 = 3/8,$ this would be proportional to $Q_N$, 
reflecting the same group-theoretical origin of these models.

In either model, one linear combination of the three $S$ fermions 
(call it $S_3$) becomes massive by combining with the 
(neutral) gaugino from $U(1)_N$ or $SU(2)'_R$ breaking, resulting in
$m_{S_3} \simeq M_{Z'},$
with $M_{Z'}/M_Z \simeq (25s_w^2/6) (u^2/v^2) = 0.96 (u^2/v^2)$ in the former 
\cite{kema}, and $M_{Z'}/M_Z \simeq [(1-s_w^2)^2/(1-2s_w^2)] (u^2/v^2) = 1.10 
(u^2/v^2)$ in the latter \cite{bahema}, 
where $u = \langle \tilde S_3 \rangle$.  The other two $S$ fermions are 
presumably light and could be 
considered ``sterile'' neutrinos \cite{ma1,ma3}.  Hence the invisible 
width of $Z'$ is predicted to have the property
\begin{equation}
\Gamma (Z' \to \nu \bar \nu + S \bar S) = \left( {62 \over 15} \right) 
\Gamma (Z' \to l^- l^+),
\end{equation}
in the $U(1)_N$ model, and
\bea
\Gamma (Z' \to \nu \bar \nu + S \bar S) = \left( {5 -16s_w^2 + 14s_w^4 
\over 6 - 30s_w^2 + 39s_w^4} \right) \Gamma (Z' \to l^- l^+) \nn,
\eea
in the skew left-right model.

In addition to the extra neutral gauge boson $Z'$, there is also the 
charged gauge boson $W_R^\pm$ in the skew left-right model.  It has the 
unusual property that it carries nonzero $B-L$ as explained before. 
The mass  of $W_R$ is given by
\begin{equation}
M_{W_R} \simeq \left( {\cos 2 \theta_W \over \cos \theta_W} \right) M_{Z'} 
= 0.84 ~M_{Z'}.
\end{equation}
It is predicted to decay only into 2 out of the 3 charged leptons because 
$S_3$ is heavy and its partner in the $SU(2)'_R$ doublet is necessarily 
a mass eigenstate, i.e. $e^c$, $\mu^c$, or $\tau^c$.  If, for example, it 
is $\tau^c$, then $W_R^+$ may decay only into $e^+ S$ or $\mu^+ S$, but 
not to $\tau^+ S.$

The Yukawa interactions differ in
the $U(1)_N$ Models 1 and 2, and in the skew $SU(2)'_R$ model, 
as explained before.  Perhaps the most distinctive experimental signatures
in this sector are the s-channel diquark $h$ resonances at hadron 
colliders predicted in the $U(1)_N$ Model 2.  At the LHC, the initial 
state from 2 valence quarks carries $B=2/3$, hence a diquark resonance 
may occur without suppression.  This allows us to test the existence of 
the diquark $h$ above 5 TeV \cite{res}.

In conclusion, in the context of $E_6$ superstring theory,
the requirement of successful leptogenesis uniquely leads 
to only two possible extensions of the SM at the TeV energy scale:
$SU(3)_C \times SU(2)_L \times U(1)_Y \times U(1)_N$ and
$SU(3)_C \times SU(2)_L \times SU(2)'_R \times U(1)_{Y_L + Y'_R}.$ 
Two Yukawa structures are possible in the former model,
but only one in the latter.
There are more sources of leptogenesis in these models than in the
standard Fukugita-Yanagida scenario, while
the smallness of Majorana neutrino masses is assured by the standard seesaw 
mechanism. They are also 
the only two such extensions of the SM which are within the 
1$\sigma$ contour of present neutral-current data.
This fact allows for the exciting possibility of discovering 
the extra $Z'$ boson with the predicted couplings in either model, 
the unusual $W_R^\pm$ boson in the skew left-right model,
and the diquark resonances in the $U(1)_N$ Model 2.

This work was supported in part by the U.~S.~Department of Energy under Grant 
No.~DE-FG03-94ER40837. T.H. and U.S. acknowledge the 
Physics Department, University of California at Riverside for hospitality.

\bibliographystyle{unsrt}

\begin{thebibliography}{99}
\bibitem{ft}
Baryogenesis from sphalerons directly (without any preexisting lepton 
asymmetry) is strongly constrained by experimental bounds on the Higgs-boson 
and superparticle masses. We do not consider this possibility here.
\end{thebibliography}

\end{document}